\documentclass[aps,prc,preprint,tightenlines,
               superscriptaddress,byrevtex]{revtex4}

\newcommand{\NN}{$N\!N$\,}

\begin{document}

\preprint{THEF-NIJM 04.05}

\title{Partial-Wave Analyses of all Proton-Proton and Neutron-Proton
       Data Below 500 MeV%
       \footnote{Oral presentation at the 19th European Conference on
                 Few-Body Problems in Physics, Groningen, The Netherlands,
                 23--27 August 2004. \\
                 To be published in the proceedings.}
      }

\author{M.C.M. Rentmeester}
\affiliation{Institute for Theoretical Physics,
             Radboud University Nijmegen,
             Nijmegen, The Netherlands}

\author{R.G.E. Timmermans}
\affiliation{Theory Group,
             KVI,
             University of Groningen,
             Groningen, The Netherlands}

\author{J.J. de Swart}
\affiliation{Institute for Theoretical Physics,
             Radboud University Nijmegen,
             Nijmegen, The Netherlands}

\begin{abstract}
In 1993 the Nijmegen group published the results of energy-dependent
partial-wave analyses (PWAs) of the nucleon-nucleon (\NN) scattering
data for laboratory  kinetic energies below $T_{lab}=350$
MeV (PWA93)\cite{1}.
In this talk some general aspects, but also
the newest developments on the Nijmegen \NN PWAs
are reported.
We have almost finished a new energy-dependent PWA and will discuss
some typical aspects
of this new PWA; where it differs from PWA93, but also what
future developments might be, or should be.
\end{abstract}

\maketitle

In order to learn about the \NN interaction scattering
experiments have been the primary source of experimental
information for well over half a century. A substantial
database of $pp$ as well as $np$ scattering
data has been established.
The available data cover a large range of energies, angles,
and also different observables, see, e.g., Ref.\cite{2}.
The big question however is, what is it that we can learn
about the \NN interaction from all these data, and, how
can we extract that knowledge from the data.

The \NN interaction is far from trivial. We are dealing with
the a strong as well as electromagnetic interaction, and
nucleons are spin-half particles as well as isospin-half
particles. This allows for a rich and complicated structure
in the \NN interaction that needs to get unraveled.
Not only are we interested in a qualitatative description
of the \NN interaction and its characteristic properties,
we also wish to establish a quantitatively sound description
of the \NN interaction.

The tool of choice to learn about the \NN interaction has
since day and age been the partial-wave analysis (PWA) of
the data.
In general one could describe the PWA as the tool that tries
to extract as much information as possible about the \NN
interaction from the experimental scattering data (values
plus uncertainties) in a preferably model-independent way,
and express that information in phase shifts (again, values
and uncertainties) and other relevant quantities, e.g.
coupling constants.

Of course this definition is very general definition. It
allows for various approaches to implement a PWA, and
subsequently various \NN PWAs do exist, see, e.g.,
Refs.~\cite{1,3a,3b}.

In the Nijmegen energy-dependent \NN PWA we try to exploit
as much as possible what is ``known'' about the \NN
interaction and parametrize what is ``unknown'' in a
phenomenological way. We have chosen to do so by using
a boundary condition approach. For each partial wave
the relativistic Schr\"odinger
equation is solved for $r > b$, with the
``known'' long-range and intermediate-range interaction
(strong and electromagnetic). For $r = b$ we introduce an
energy-dependent boundary condition representing a
phenomenological parametrization of the ``unknown''
short-range physics.

In 1993 we established this way an energy-dependent PWA
of all $pp$ and $np$ data below $T_{lab} = 350$ MeV \cite{1}.
We have now almost finished a new PWA, up to 500 MeV.
Below we will remark on some of the typical aspects
of this new PWA; where it differs from PWA93, but also what
future developments might be, or should be.

\begin{itemize}
\setlength{\itemsep}{\baselineskip}

\item
The energy range of the PWA has been extended
from 350 MeV to 500 MeV.
This means that we have gone well above the pion-production
threshold and consequently inelasticies start to become
significant. We introduce them in selected waves by employing
a complex energy-dependent boundary condition ar $r = b$.
While this gives satisfactory results for now,
a more physical guidance by allowing for physical mechanisms
that generate inelasticities might be better. Not only for the
waves where we could not introduce inelasticities now, but also
if we wish to extend our PWA to even higher energies.

\item
The \NN database has been enlarged. Not only with the data
between 350 and 500 MeV, but also with many new data below
350 MeV\cite{2}. Currently we use approximately 5000 $pp$
and 5000 $np$
data below 500 MeV. These numbers sound impressive and they do
allow us to perform a PWA. But quantity does not equal quality.
For example, the $np$ data are much less accurate and varied than
the $pp$ data, and often suffer from systematic errors\cite{4}.
Many data, especially older, but also (very) recent data, are
of mediocre quality. More and better data are needed in order
to produce more reliable results. Also if we want to determine
more precise values for coupling constants, or if we want to
perform a quantitative study of small effects such as charge
independence breaking.

\item
We have been able to extract uniquely all the important $np$
phase parameters, $I=1$ as well as $I=0$, from the $np$ data.
Such a separate PWA of the $np$ data$^3$, without any input
from the $I=1$ $pp$ partial waves, is less model dependent,
and, moreover, a comparison of the phase parameters from such
an independent $np$ PWA with the corresponding ones obtained
from the $pp$ data provides information about possible
charge-independence breaking in the $I=1$ \NN waves.

\item
In PWA93 we used for the long-range interaction a modified
version of the Nijm78 potential\cite{4a}. This has now been replaced by
the One-Pion Exchange potential, and a chiral Two-Pion Exchange
potential.
As this long-range interaction is the result of a systematic
expansion, it is, in
principle, model independent and could be improved, if necessary,
by taking higher orders into account. Using this long-range
\NN interaction we obtained a better fit to the data
\cite{5}. Moreover we were able to determine values for
some of the low-energy coupling constants \cite{6}.

\item
$np$ differential cross sections are notoriously difficult
to measure. Hence the large number of sets of data that
differ significantly from each other. Some, or even many of
them, suffer from systematic errors. While we cannot deal with
systematic errors in general, certain systematic uncertainties,
however, can be
treated. By taking the effects of relative normalization
procedures into account we got an improved fit
to the data\cite{4}. Some data that we previously had to omit
can now be included in a fit.
Systematic effects will
play a bigger role when experiments get more complicated and
statistics get better. Properly handling systematic
uncertainties in analyses will become more important. As that
requires insight in experimental procedures, more
cooperation with experimentalists is imperative.

\end{itemize}

We are
currently in the exploratory phase, i.e. carefully studying
the results of the new PWA and do some finetuning where deemed
necessary. More details of the analyses and its results will
be published in due time.

\end{document}